\documentclass[aps,prb,reprint,showpacs,showkeys,superscriptaddress]{revtex4-1}
\usepackage{multirow}
\usepackage{graphicx}
\usepackage{bm}
\usepackage{dcolumn}
\usepackage{picture}
\usepackage[hidelinks]{hyperref}
\usepackage{natbib}
\usepackage{amsmath}
\usepackage{times}
\begin{document}
\title{Impurities in weakly coupled quantum spin chains  Sr$_{2}$CuO$_{3}$ and SrCuO$_{2}$}

\author{Koushik Karmakar} \affiliation{Department of Physics, Indian Institute of Science Education and Research, Pune, Maharashtra-411008, India}

\author{Rabindranath Bag}\affiliation{Department of Physics, Indian Institute of Science Education and Research, Pune, Maharashtra-411008, India}

\author{Markos Skoulatos}\affiliation{Laboratory for Neutron Scattering and Imaging, Paul Scherrer Institute, 5232 Villigen, Switzerland}\affiliation{Heinz Maier-Leibnitz Zentrum (MLZ) and Physics Department E21, Technische Universität München, D-85748 Garching, Germany}

\author{Christian R\"uegg}\affiliation{Laboratory for Neutron Scattering and Imaging, Paul Scherrer Institute, 5232 Villigen, Switzerland}\affiliation{Department of Quantum Matter Physics, University of Geneva, 1211 Geneva, Switzerland}

\author{Surjeet Singh}
\email[email:]{surjeet.singh@iiserpune.ac.in}
\homepage[url:]{http://www.iiserpune.ac.in/~surjeet.singh}
\affiliation{Department of Physics, Indian Institute of Science Education and Research, Pune, Maharashtra-411008, India}\affiliation{Center for Energy Science, Indian Institute of Science Education and Research, Pune, Maharashtra-411008, India}

\date{\today}

\begin{abstract} We study the effect of non-magnetic Zn$^{2+}$ (spin-$0$) and magnetic Ni$^{2+}$ (spin-$1$) impurities on the ground state and low-lying excitations of the quasi-one-dimensional spin-$1/2$ Heisenberg antiferromagnet Sr$_{2}$CuO$_{3}$ using inelastic neutron scattering, specific heat and bulk magnetization measurements. We show that 1 \% Ni$^{2+}$  doping in Sr$_2$CuO$_3$ results in a sizable spin gap in the spinon excitations, analogous to the case of Ni doped SrCuO$_2$ previously reported [ref. \citenum{Simutis2013}]. However, a similar level of Zn$^{2+}$ doping in SrCuO$_2$, investigated here for comparison, did not reveal any signs of a spin gap. Magnetic ordering temperature was found to be suppressed in the presence of both Zn$^{2+}$ and Ni$^{2+}$ impurities, however, the  rate of suppression due to Ni$^{2+}$  was found to be much more pronounced than for Zn$^{2+}$. Effect of magnetic field on the ordering temperature is investigated. We found that with increasing magnetic field, not only the magnetic ordering temperature gradually increases but the size of specific heat anomaly associated with the magnetic ordering also progressively enhances, which can be qualitatively understood as due to the field induced suppression of quantum fluctuations.  \end{abstract}


\maketitle

\section{Introduction}

Antiferromagnetic spin-$1/2$ chains are model systems for realizing a wide range of interesting quantum many-body ground states. In particular, the ground state of spin-$1/2$ Heisenberg antiferromagnetic chain (HAFC) is characterized as a quantum critical spin liquid in which the spin correlations decay as a power-law \cite{Giamarchi2003}. However, despite the absence of static long-range order, it manifests well-defined spin-$1/2$ excitations called spinons. The spinons are created in pairs leading to a quantum continuum of gapless two-spinons states \cite{Faddeev1981}. In systems where the spin-$1/2$ HAFC can be realized, presence of interchain coupling, disorder, spin frustration and/or applied magnetic field bring about novel and unexpected changes to the low-energy properties \cite{Alloul2009, Zheludev2013}. In the last few decades, quantum field theories have been successfully employed in investigating some of these properties and in predicting new phenomena \cite{Fradkin2013}. Of particular interest are the field theory results for the spin-$1/2$ HAFC in the presence of magnetic and non-magnetic impurities [for review see, ref. \citenum{Affleck1995}]. Depending upon the size of the impurity spin and its exchange coupling with the host spins several interesting scenarios, including multi-channel Kondo effect and non-fermi liquid behaviors are expected to arise \cite{Eggert1992, Eggert2001}. In the specific case of an antiferromagnetically coupled spin-$1$ impurity, theory predicts a Kondo-singlet ground state where the impurity spin is Kondo-screened by the two neighboring spins of the chain, resulting in an open chain with three sites removed.  

These arguments can be extended to dilute concentration of spin-$1$ impurities in the chain. For low doping concentrations, the ground state consist of Kondo-singlets at each impurity site, breaking the periodic chain into finite-length sections. Such a finite-size confinement is expected to open up a spin gap in the low-lying spin excitations of the doped chain. Recently, it was shown that merely 1 \% of Ni doping in SrCuO$_2$ opens-up a sizeable spin pseudogap of about 8 meV, not related to any structural transition \cite{Simutis2013}. It was proposed that Ni spin (spin-1) in SrCuO$_2$ is Kondo screened which causes the spin gap to open. 

Since a spin-0 impurity in the spin-$1/2$ chain also disrupts the translational invariance of the chain, it is pertinent to ask if an analogous spin-gap will also appear in the presence of a spin-0 impurity. We experimentally investigate this issue by doping SrCuO$_2$ with dilute concentration of Zn$^{2+}$ (spin-0) impurities. Similarly, it is also important to investigate if an equivalent spin gap will result when Ni is doped in spin-$1/2$ chains other than SrCuO$_2$, which will add further credence to the theory. Finally, the effect of spin gap on the magnetic and thermodynamic properties of weakly coupled quantum spin chains has not been properly explored . Here, we investigate these issues; and also the effect of magnetic field on the ordering temperature of pure and doped spin-$1/2$ chains. 
The model systems employed in our investigations are the quasi-one-dimensional spin-$1/2$  HAFC compounds Sr$_2$CuO$_3$ and SrCuO$_2$. These are Mott insulators with an exceptionally good one-dimensionality because of  a very small value of inter- to intrachain coupling ratio ($< 10^{-3}$) [ref. \citenum{Motoyama1996}]. In the past, both these compounds have been extensively used in establishing several fundamental predictions of the spin-$1/2$ HAFC model [see, for example, Refs. \citenum{Motoyama1996, Zaliznyak2004, Kim2006, Schlappa2012}]. 

We show here that Ni-doping in Sr$_2$CuO$_3$ indeed results in a spin gap analogous to the Ni-doped SrCuO$_2$. Similar level of Zn-doping in SrCuO$_2$, however, does not produce a spin gap. We further show that  Ni- and Zn-doping in Sr$_2$CuO$_3$ suppress the long-range magnetic ordering temperature. However, in the Ni-case the suppression is shown to be much more pronounced than expected from the quantum chain mean-field theory. We also show that with increasing magnetic field the transition temperature of undoped and 1 \% Zn-doped Sr$_2$CuO$_3$ shows a gradual isotropic increase (at 1.5 \% per Tesla); concomitantly, the specific heat anomaly become more pronounced in agreement with the theoretical expectation for weakly-coupled quantum spin chains.      

The outline of our paper is as follows: experimental details are presented in section II  followed by results and discussion in section III. Summary of the results obtained and conclusions drawn appear in the last section (IV). Section III is further divided into four subsections as follows: III-A deals with inelastic neutron scattering. This is followed by magnetic ordering suppression due to Zn- and Ni-doping in III-B, effect of external field are presented in III-C; and Schottky contribution in the specific heat due to odd-length chain segments is discussed in subsection III-D.    

\section{Experimental Details}

Single crystals of Sr$_2$CuO$_3$, Sr$_2$Cu$_{0.99}$Ni$_{0.01}$O$_3$, Sr$_2$Cu$_{0.99}$Zn$_{0.01}$O$_3$ and SrCu$_{0.98}$Zn$_{0.02}$O$_2$ were grown using the travelling-solvent floating-zone (TSFZ) technique in an infrared image-furnace (Crystal System Corporation, Japan). The polycrystalline feed rods for the crystal growth experiments were synthesized using high-purity precursors ($> 99.99 \%$). The crucible-free floating zone technique, and the absence of any foreign flux or solvent during  the TSFZ experiments ensured that unintended impurities, either from the crucible material or from the flux, are not incorporated in the grown crystals. This point is crucial to our experiments since the purpose of this study is to explore the consequences of  a specific impurity type on the physical properties. Details of crystal growth and compositional/structural characterizations have been previously reported in Refs. [\citenum{Karmakar2014, Karmakar2015b}]. 

\begin{figure}\centering \includegraphics[width=8.5 cm]{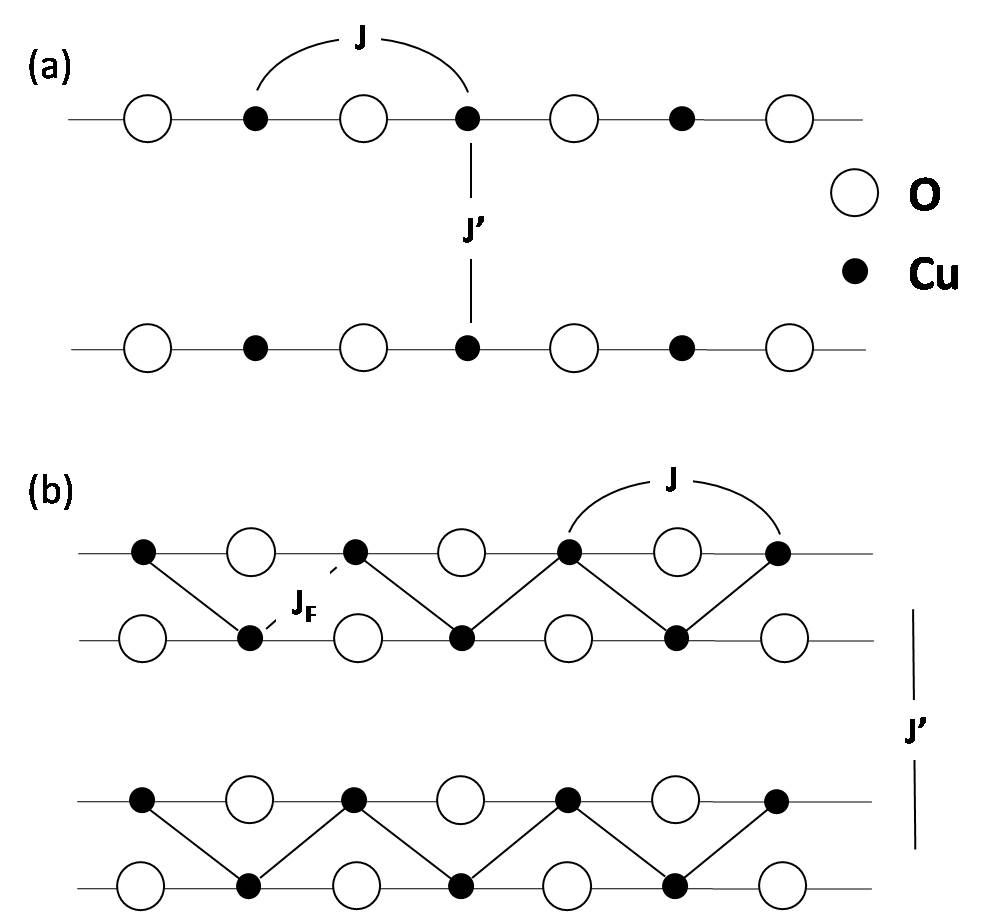} \caption{Schematics showing (a) linear  -Cu-O-Cu- chain in Sr$_2$CuO$_3$, and (b) zigzgag chain in   SrCuO$_2$. J and J' are intra- and inter-chain interactions respectively; J$_F$ represents the weak ferromagnetic coupling within the zigzag chain (see text for details)} \label{CuOCu} \end{figure}

All the grown crystals were found to crystallize with the expected orthorhombic structures previously reported for Sr$_2$CuO$_3$ and SrCuO$_2$ \cite{Chr1971}. No signs of extra-phases crystallizing along with the main phase could be detected either in the electron microscopy or x-ray diffraction experiments. The structure of Sr$_2$CuO$_3$ consists of linear -Cu-O-Cu- chains running parallel to the $b$-axis. In the case of SrCuO$_2$, the chains are zigzag shaped and they are oriented parallel to the $c$-axis of the unit cell. The zigzag chain can be visualized as a two-leg ladder where every Cu-ion in one leg has an O-ion directly across it in the other and vice versa (shown schematically in Fig. \ref{CuOCu}) . The Cu$^{2+}$ moments within a leg couple antiferromagnetically via the $180^{\circ}$ Cu-O-Cu superexchange path (J $\approx$ $2000$ K). Additionally, the legs are weakly coupled via a nearly $90^{\circ}$ ferromagnetic Cu-O-Cu superexchange path (J$_F \approx$ 0.1- 0.2 J) \cite{Motoyama1996}, which is known to frustrate the main antiferromagnetic exchange J  (Ref. \citenum{Zaliznyak1999}). Both compounds have similar  intrachain J and interchain J$'$ ($\approx$ 10$^{-3}$ J) couplings \cite{Motoyama1996}. 

The concentration of Zn in the grown crystals was estimated to be significantly smaller than the nominal value \cite{Karmakar2015b}. In Sr$_{2}$Cu$_{0.99}$Zn$_{0.01}$O$_{3}$, for example, using the susceptibility analysis it was estimated to be $\sim0.006$ per Cu \cite{Karmakar2015a}. An independent estimation of Zn concentration in Sr$_{2}$Cu$_{0.99}$Zn$_{0.01}$O$_{3}$ and $_{0.98}$Zn$_{0.02}$O$_2$ was made by Yannic Utz et al. \cite{Yannic2017} using the Inductively Coupled Plasma (ICP) technique. They found these to be $\sim 0.0036$ and $\sim 0.015$, respectively. In the Ni-doped crystal, however, the actual dopant concentration was found to be in good agreement with the nominal value. Hereafter, in this paper we shall use only the nominal composition. 

The crystals were oriented using a Laue camera (Photonic Science, UK \& France). The oriented crystal pieces were annealed under Ar flow for 72 hours at $900^{\circ}$ C before measurements to minimize additional oxygen defects. INS experiments were carried out on oriented single crystals of mass $ m \approx2$ g using the thermal triple-axis spectrometer EIGER \cite{Stuhr2017} at SINQ, Paul Scherrer Institute, Switzerland. The final neutron energy of the spectrometer was set to $E_{f}=14.7$ meV, which was filtered using a graphite filter. Specific heat and magnetization measurements were performed using calorimeter and vibrating sample magnetometer attachments in a physical property measurement system (Quantum Design, USA).

\section{Results and Discussion}

\subsection{Low-lying excitations of the doped chains} 
\begin{figure}\includegraphics[width=8.5cm]{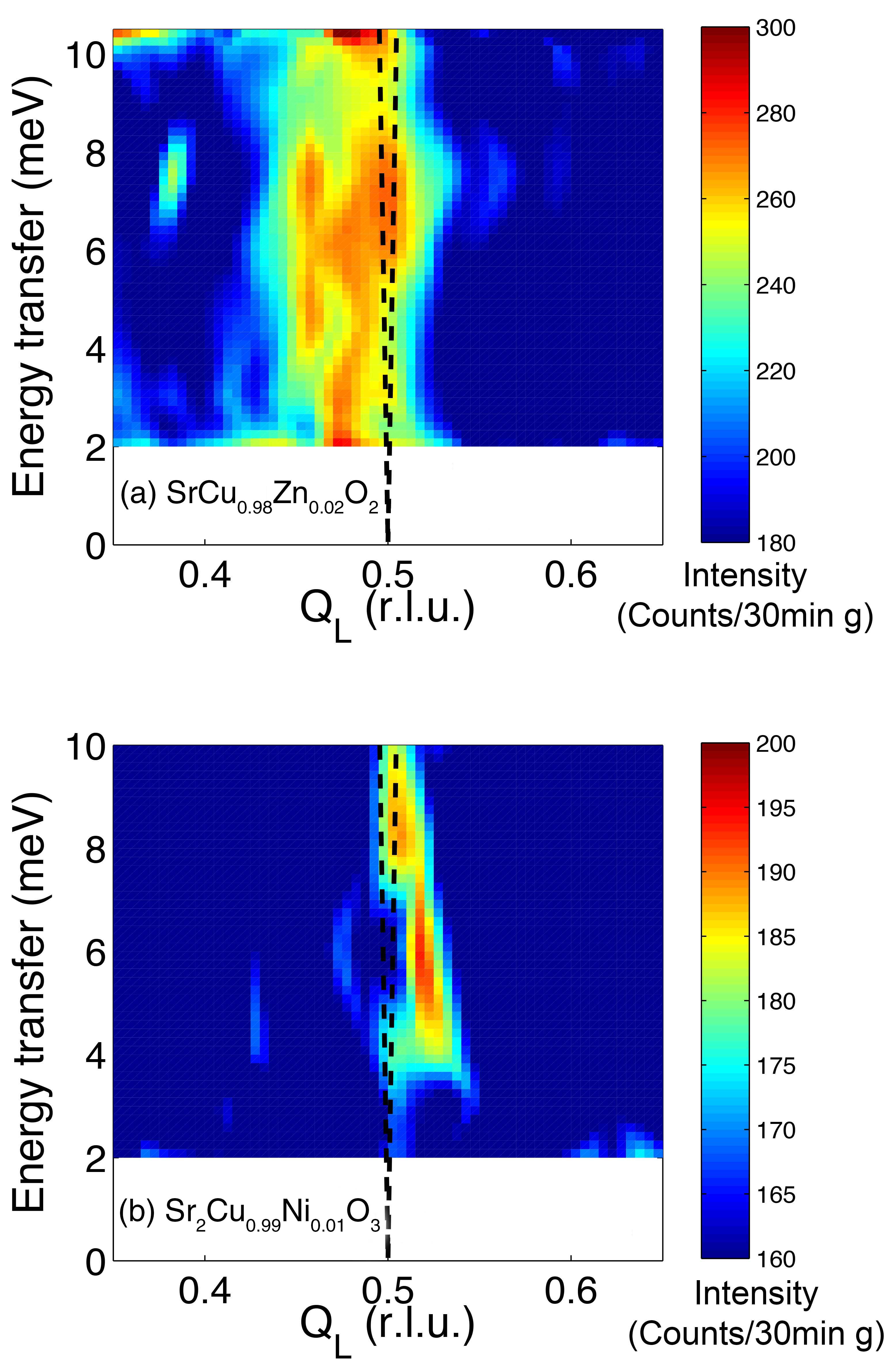} \caption{Triple-axis neutron spectra constructed by constant energy transfer scans  (2-10 meV) for (a) SrCu$_{0.98}$Zn$_{0.02}$O$_{2}$, and (b) Sr$_{2}$Cu$_{0.99}$Ni$_{0.01}$O$_{3}$.} \label{INS} \end{figure}

Inelastic neutron scattering (INS) data of a SrCu$_{0.98}$Zn$_{0.02}$O$_2$ single-crystal is shown in Fig. \ref{INS}(a). We focus here on the spinon dispersion for energy transfer below 10 meV around the point Q = (1/2 0 1/2). The low-energy excitation spectrum is shown by an intensity map, constructed from individual scans measuring the cross-section S(Q, $\omega$) along constant energy transfers. Due to the weak scattering signal each point is measured for 30 minutes. The vertical dashed lines at Q$_L$ = 0.5 r.l.u. represent the calculated lower-bound of the two-spinon continuum for J = 230 meV \cite{DesCloizeaux1962}. 

As shown in Fig. \ref{INS}(a), Zn-doping causes only minor variations of the INS intensity as a function of energy transfer down to the instrumental energy resolution of 2 meV. In particular, we do not see INS intensity disappearing at low-energy transfers indicating that the excitation spectrum is gapless down to 2 meV. This should be contrasted with the case of SrCu$_{0.99}$Ni$_{0.01}$O$_2$ crystal reported recently \cite{Simutis2013}, where the INS intensity gradually disappeared at low-energy transfers due to opening of a spin pseudogap. It is interesting that while Ni-doping in SrCuO$_2$ opens-up a gap, similar level of Zn doping does not. 
To investigate if gapping upon Ni-doping is specific to SrCuO$_2$ we measured INS spectrum of the crystal Sr$_{2}$Cu$_{0.99}$Ni$_{0.01}$O$_3$. The measured INS spectrum around the point Q = (0 1/2  1/2) is shown in Fig. \ref{INS}(b). These measurements were done under the same conditions as in Fig. \ref{INS}(a). The INS intensity in this case indeed became very weak below an energy transfer of $\approx 4$ meV , indicating the presence of a spin gap in the excitation spectrum. The momentum-integrated intensity  S($\omega$) = $\int$S(Q, $\omega$) dQ was also found to approximately fall on the theoretical S($\omega$) calculated for 1 \% of randomly positioned defects using the envelope function derived in Ref. \citenum{Simutis2013}.

Zn$^{2+}$ ion is characterized by its completely filled d-shell (\textit{d}$^{\textit{10}}$) that makes it a magnetically inactive spin-0 ion. Therefore, Zn$^{2+}$ when doped in the chains is expected to break them into finite-length segments. Ni$^{2+}$ (\textit{d}$^{\textit{8}}$), on the other hand, can either be in a low-spin (spin-$0$) or high-spin (spin-$1$) state. Simutis et al. in their paper assumed that the doped Ni$^{2+}$ is in the high-spin state \cite{Simutis2013}. They argued that the gapping is a consequence of the finite-size effects due to Kondo screening of the Ni$^{2+}$ spin as proposed theoretically \cite{Eggert1992}. In this picture, the Kondo cloud at the impurity site can be regarded as an extended three sites long  spin-$0$ defect, which magnetically isolates the chain segments on either side of it. Confinement of spinons over these finite-size segments results in the spin gap whose size scales inversely with the average length of the chain segments. Since the impurities are statistically distributed in the chain, the gap is expected to be soft or, in other words, a pseudogap. Recent NMR experiments, probing the spin-lattice relaxation rates in Ni-doped chains of varying Ni-concentrations confirmed this hypothesis  \cite{Utz2015}. In contrast to this, in the presence of a Zn$^{2+}$ impurity, the chain segments remain weakly bridged via the second-nearest-neighbour (\textit{nnn}) interaction (J$_{nnn}$ $\approx$ 140 K, see Ref. \citenum{Rosner1982}) leading only to a weak quasiparticle confinement, which is probably not enough to open a measurable gap.

\begin{table*} \caption{Values of fitting parameters in the specific heat of Pure, Zn and Ni doped crystals}\label{Para_Cp}
	\begin{ruledtabular}
		\begin{tabular}{l c c c c c}
			\multirow{2}{*}{Parameters} & $\alpha$ & $\beta$ & $\delta$ & $J$ & $\Theta_{D}$ \\
			& (mJ/mol.K$^{-2}$) & (mJ/mol.K$^{-4}$) & (mJ/mol.K$^{-6}$)& (K) & (K) \\
			\hline
			Sr$_{2}$CuO$_{3}$ & 2.53(2) & 0.130(1) & 5.7$\times 10^{-5}$& 2198 $\pm$ 100 & 448 \\
			Sr$_{2}$Cu$_{0.99}$Zn$_{0.01}$O$_{3}$ & 2.65(2) & 0.136(1) & 4.7$\times 10^{-5}$& 2090 $\pm$ 100 & 441 \\
			Sr$_{2}$Cu$_{0.99}$Ni$_{0.01}$O$_{3}$ & 0.164(8) & 0.134(1) & 3.5$\times 10^{-5}$ & -- & 443 \\
		\end{tabular}
	\end{ruledtabular}
\end{table*}

An alternative view is presented in Ref.[\citenum{Yannic2017}]. By comparing the NMR spectra of Zn$^{2+}$ and Ni$^{2+}$ doped chains, it has been argued that the doped Ni$^{2+}$ ions in these compounds are in their low-spin state. In the square-planar coordination, the crystal fields can be strong enough to favor a low-spin state \cite{Wang}. However, if true then one may argue that since the low-spin Ni$^{2+}$ is magnetically equivalent to Zn$^{2+}$, doping with Zn$^{2+}$ should also produce a spin gap analogous to doping with Ni$^{2+}$. However, no such gap has been found in our experiment on the Zn-doped crystal down to 2 meV energy transfer; even though the concentration of Zn in SrCu$_{0.98}$Zn$_{0.02}$O$_{2}$ is more than that of Ni in Sr$_{2}$Cu$_{0.99}$Ni$_{0.01}$O$_{3}$. This can only be understood if one assumes that the concentration of Zn entering the chain is smaller than the actual Zn concentration in the crystal measured using the ICP technique in ref. \citenum{Yannic2017}. In which case the remaining Zn in the crystal would either go into the interstitial sites and/or substitutes Sr$^{2+}$ in the structure. However, ionic radius of Zn$^{2+}$ at {0.6 \AA} is too large to easily fit at the interstitial site and too small to replace Sr$^{2+}$ whose ionic radius is close to {1.2 \AA}. From these considerations it appears unlikely that Ni$^{2+}$ is in a low-spin state. Further experiments to ascertain the spin-state of Ni in these compounds may be helpful in gaining clarity on this point.

Before closing this section we should mention that spin gaps were also previously reported in 10 \% Ca$^{2+}$ doped Sr$_{2}$CuO$_{3}$ \cite{HammerathPRB} and SrCuO$_{2}$ \cite{HammerathPRL}. Since Ca replaces Sr outside the chains leaving them unbroken or continuous, it can therefore be inferred that gapping of excitation spectrum can also be due to reasons other than the finite-size effects discussed above . However, when doped outside the chain, the doping concentration required to produce a comparable gap is significantly high. Interestingly, not every impurity type doped in the chain leads to a spin gap. Very recently, it was shown that Co (spin-1/2) impurities in SrCuO$_{2}$ do not alter the gapless nature of its spin excitations \cite{Karmakar2017a}.    

\begin{figure}\centering \includegraphics[scale=1]{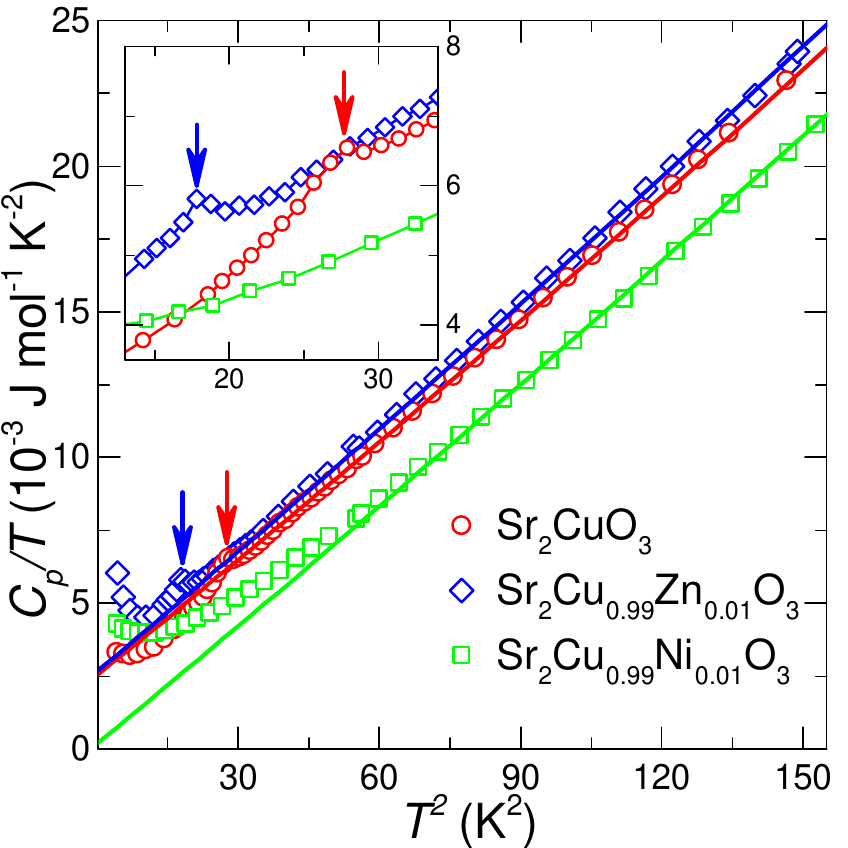} \caption{$C_{p}/T$ plotted against $T^2$ for Sr$_2$CuO$_3$, Sr$_2$Cu$_{0.99}$Zn$_{0.01}$O$_3$ and Sr$_2$Cu$_{0.99}$Ni$_{0.01}$O$_3$. The solid lines are fit to the data using equation \ref{Spheat}. Inset: an expanded view of the data shown in the main panel to highlight the anomalies associated with the magnetic ordering marked using arrows.} \label{ZF_Cp} \end{figure}

\subsection{Magnetic ordering temperature of the doped chains} In Fig. \ref{ZF_Cp}, temperature variation of specific heat (C$_p$) for the undoped and doped Sr$_2$CuO$_3$ crystals is shown from T = 2 K to 15 K. The low-temperature data are shown in the inset. C$_p (T)$ of undoped and Zn-doped crystals show distinct anomalies at T = 5.4 K and 4.3 K, respectively. These anomalies correspond to the long-range ordering of the Cu-moments. Sizes of these anomalies are smaller than what one expects from the ordering of a full spin-1/2 moment because the ordered Cu-moment in the spin-$1/2$ chains is highly reduced $(\sim0.06 \mu_B)$  due to the strong quantum fluctuations \cite{Kojima1997}. Interestingly, C$_p (T)$ of Ni-doped crystal, shown in Fig. \ref{ZF_Cp}, varies smoothly over the whole temperature range showing no discernible anomaly, suggesting absence of long-range order at least down to T = 2 K. Specific heat of Ni-doped crystal is also found to be smaller than that of the undoped and Zn-doped crystals. Both these observations are  consistent with the presence of a spin gap shown using the INS experiments in section IIIA. 

Above the ordering temperature, the measured specific heat in these electrically insulating crystals has principal contributions due to phonons and magnetic excitations. Therefore, the total specific heat above T$_N$ (for T $\ll \Theta_{D}$) can be written as: 

\begin{equation}\label{Spheat}
C_{p} (T) =\frac{12 xN_{A}\pi^{4}k_{B}}{5}\left(\frac{T}{\Theta_{D}}\right)^{3} + \frac{2yN_{A}k_{B}}{3}\left(\frac{T}{J}\right)
\end{equation}
where $N_{A}$ is Avogadro's number and $k_{B}$ is Boltzmann's constant. $\theta$$_D$ represents the Debye temperature, and J the intrachain coupling. The first and second terms correspond to the phonon and the spinon contributions, respectively. x and y in these terms represent total number of atoms (first term) and magnetic atoms (second term) per formula unit.  By fitting the measured specific heat above T$_N$ using this expression, the values of J and $\theta$$_D$ can be obtained. The measured specific heats were fitted using the expression C$_p$ /T = $\alpha$ + $\beta$T$^2$+ $\delta$ T$^4$ in the temperature range 6 K $\leq$ T $\leq$ 12 K for the undoped crystal, and 5 K $\leq$ T  $\leq$ 11 K for the Zn  doped crystal. For the Ni-doped crystal, C$_p$ /T exhibits a gradual upturn when cooled below T = 8 K, we, therefore, fitted the data only in the linear region above this temperature.  A similar upturn is also witnessed in the specific heats of Pure and Zn-doped crystals below their respective ordering temperatures. The exact reason for this gradual low-temperature increase is not clear. It might be that there are additional magnetic degrees of freedom possibly due to uncompensated spins at the ends of the fragmented chain segments. The coefficients $\alpha$ and $\beta$ can be identified with the coefficients of the linear (spinon) and the cubic (phonon) terms in equation $( \ref{Spheat})$. The quartic term with coefficient $\delta$ is added to allow for phonon anharmonicity, which is expected to be small in this temperature range. The values of these parameters are shown in Table \ref{Para_Cp} along with the extracted values of J and $\theta$$_D$. 

Our estimation of the Debye temperature ($\approx  445 K $) agrees well with previous reports \cite{Kawamata2008,Sologubenko2000}. The  fitted value of $J$ for the undoped and Zn-doped crystals ($\approx 2000 K $) is also in good agreement with previous reports \cite{Ami1995, Eggert1996, Motoyama1996}, and also to our estimation of $J$ using the bulk magnetic susceptibility data \cite{Karmakar2015a}. Importantly, it should be noticed that for the Ni-doped crystal the fitted value of $\alpha$ is considerably smaller (Table \ref{Para_Cp}), which is consistent with the gapping of the spinon dispersion. As expected, the value of  $\delta$ in each case is very small.  

We will now briefly discuss the suppression of T$_N$ in the doped chains. By treating J' at the mean-field level, the N\'{e}el temperature (T$_N$) for weakly coupled chains, ignoring quantum fluctuations, scales as: $k_{B} T_{N} \approx zJ'(\xi_{chain}/c)$, where $\xi_{chain}$ is the correlation length within the chain at T$_N$ and \textit{c} is the lattice spacing in the chain \cite{Hone1975,Schouten1980}. From this expression it is evident that for a chain consisting of a finite number of chain breaks, the correlations will be disrupted, decreasing $\xi_{chain}$, which qualitatively explains why the N\'eel temperature decreases upon non-magnetic doping. It is worth mentioning here that unlike spin-0 impurities that suppress the transition temperature, Co (spin-1/2) impurities surprisingly increase it significantly \cite{Karmakar2017a}, which cannot be explained by these theories. A more rigorous quantum mean-field calculation of T$_N$ applicable to the spin-1/2 HAFC with non-magnetic impurities was carried out by Eggert, Affleck and Harton \cite{Eggert2002}. They found that the magnetic ordering temperature decreases sharply with increasing chain break concentration (\textit{p}) [see, Fig. 2 of ref. \citenum{Eggert2002}]). For example, a mere 0.5 \% of defect concentration is enough to suppress the ordering temperature to almost 40 \% of T$_N(0)$, where T$_N (0)$ is the ordering temperature of the hypothetical defect-less chains.

The values of \textit{p} in our crystals was estimated by a rigorous analysis of their spin susceptibility in a broad temperature range \cite{Karmakar2015a}. In the undoped crystal the value of \textit{p} was estimated to $\sim0.0047$ per Cu (i.e., $\approx 0.5 \%$). In the undoped compound these are intrinsic defects that are well-known to arise due to oxygen off-stoichiometry \cite{Motoyama1996}. For this value of \textit{p}, and the value of T$_N$ measured using specific heat, we estimate T$_N(0)$ of a hypothetical, defect-free Sr$_{2}$CuO$_{3}$ crystal to be $\sim$13 K. Since T$_N(0)$ is known, we can estimate what the magnetic ordering temperature of our doped crystals should theoretically be \cite{Eggert2002}.  For the composition Sr$_{2}$Cu$_{0.99}$Zn$_{0.01}$O$_3$  the effective $p$ value was estimated to be $\sim0.011$ per Cu, which includes chain breaks due to Zn-doping and also due to oxygen off-stoichiometry as in the undoped crystal [\citenum{Karmakar2015a}]. For this value of \textit{p} the corresponding  T$_N$ will be $\sim4.5$ K, in fair agreement with the experimental value of $4.3$ K. It should be mentioned that presence of Zn in the interstitial sites between the chains, if any, will only affect less than 1 \% of the interchain exchange pathways, which will not reduce the average interchain coupling strength substantially to account for the decrease in ordering temperature from T = 5.3 K to T = 4.3 K upon dilute Zn-doping.

\begin{figure}\centering \includegraphics[width = 8 cm]{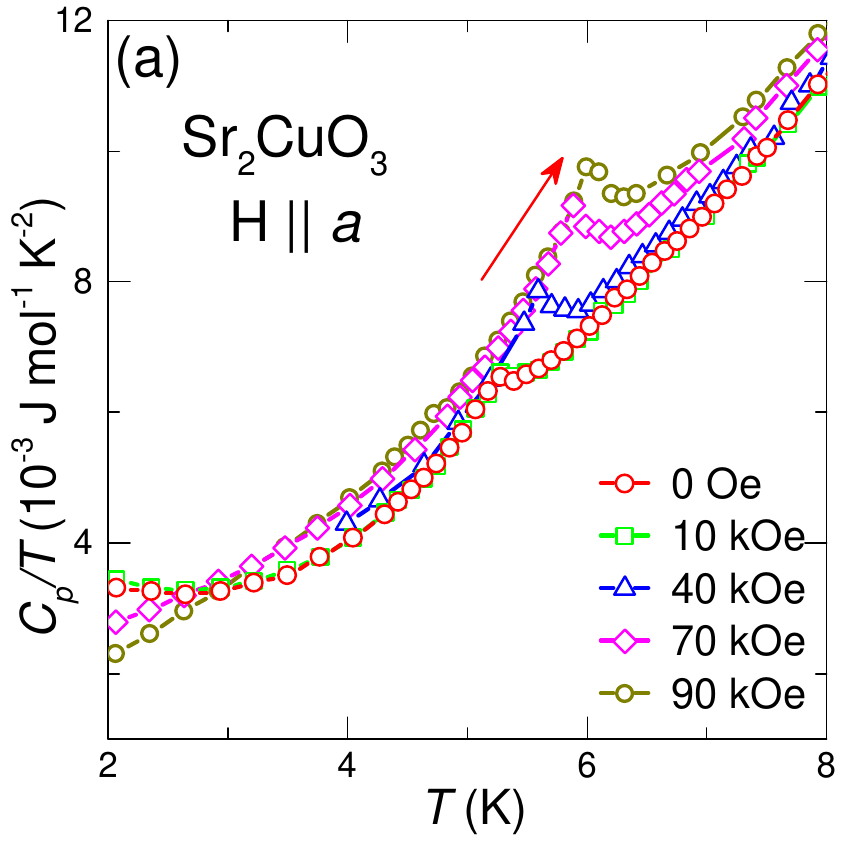}
	\includegraphics[width = 8 cm]{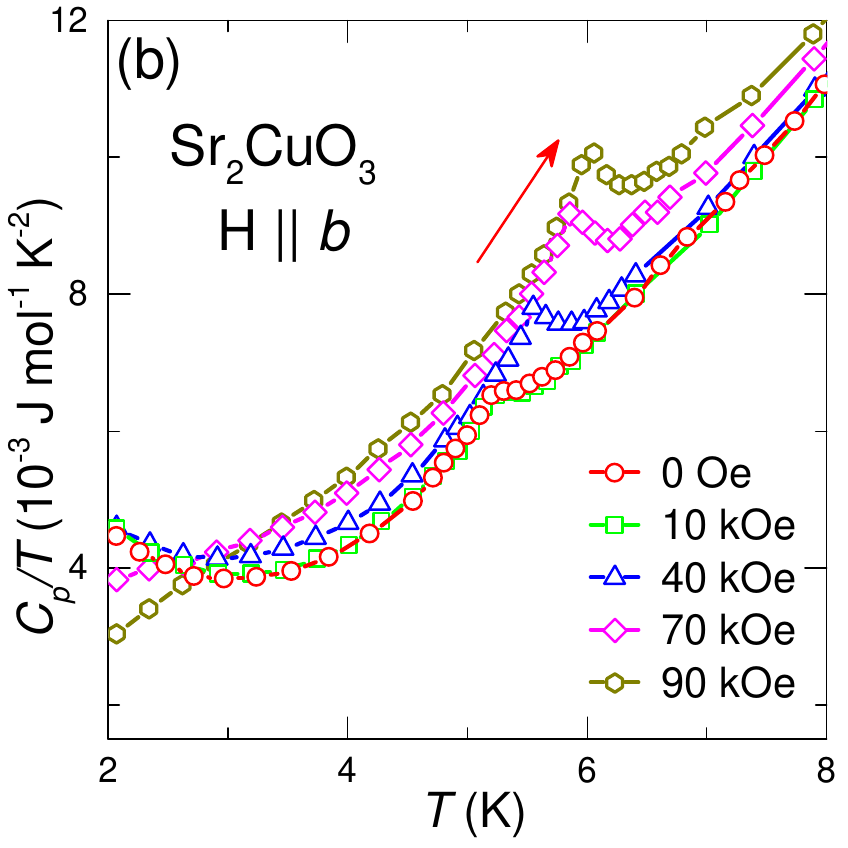} \caption{C$_p$/T of Sr$_{2}$CuO$_{3}$ plotted as a function of $T$ under various applied magnetic fields (a) $H \parallel$ \textit{a} (perpendicular to the chain). (b) $H \parallel$ \textit{b} (parallel to the chain). Arrows indicate the direction of shift of the magnetic anomaly with increasing applied field.} \label{Pure_FD} \end{figure}

Case of Ni-doping is more interesting. From the magnetic susceptibility, the effective chain break concentration in Sr$_{2}$Cu$_{0.99}$Ni$_{0.01}$O$_3$ is $\sim$0.016 per Cu, which corresponds to an expected T$_N$ of $\approx 3$ K. However, in the experimental specific heat no evidence of magnetic transition could be seen down to T = 2 K. Similar behavior is recently reported for SrCu$_{0.995}$Ni$_{0.005}$O$_2$ and SrCu$_{0.99}$Ni$_{0.01}$O$_2$ [ref. \citenum{Simutis2016}]. It was found that for low-concentration of Ni-doping ($\leq0.25 \%$) agreement between theory and experiment is acceptable but not so for higher concentrations (0.5 and 1 \% Ni) where no evidence of ordering was found down to 20 mK [\citenum{Simutis2016}]. It is therefore evident that Ni impurities suppress T$_N$ far more strongly than expected. This discrepancy could be due to the fact that magnetic excitations in Ni-doped crystals have a spin pseudogap. The agreement for low Ni-concentrations is either due to small gap size or the gap not being be fully open. $\mu_{SR}$ experiments on Ni-doped Sr$_2$CuO$_3$ crystals with progressively increasing Ni-concentration will be useful to investigate this point in the future.

\subsection{Effect of magnetic field on the magnetic ordering temperature} 
\begin{figure}\includegraphics[width=8cm]{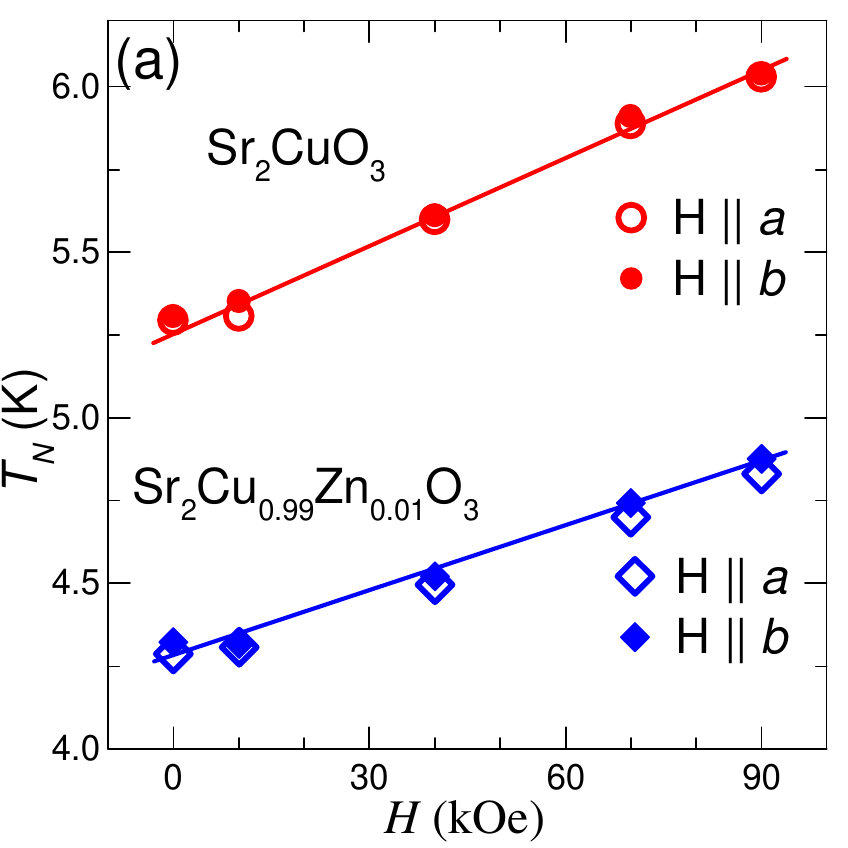}
	\includegraphics[width=8 cm]{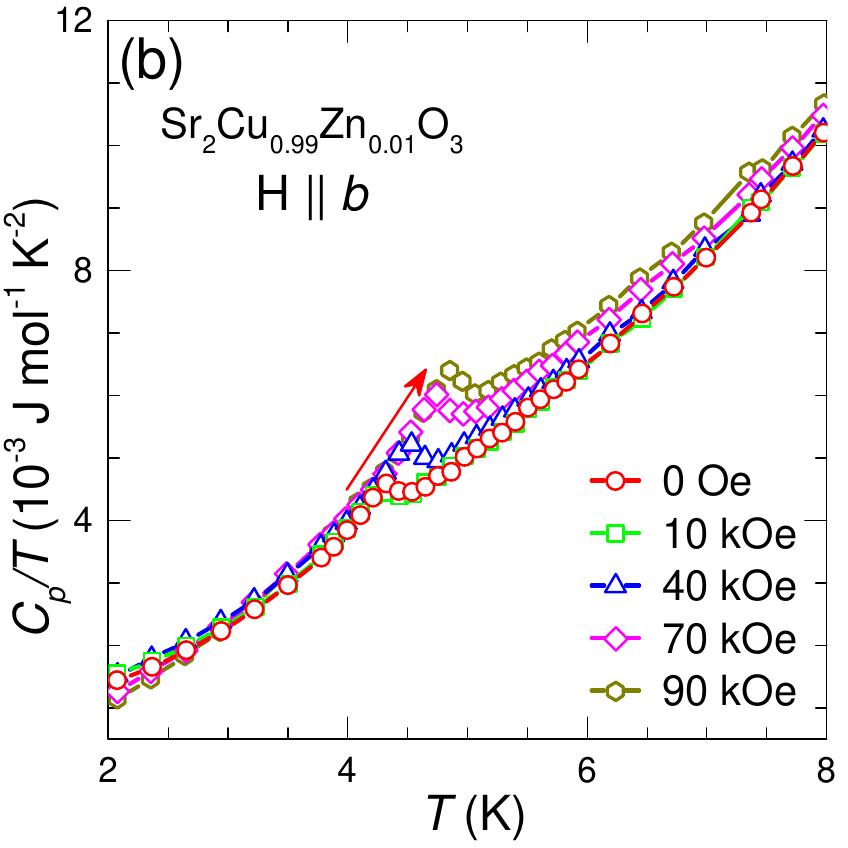} \caption{a) Variation of magnetic ordering temperature as a function of applied field in Sr$_{2}$CuO$_{3}$ and Sr$_{2}$Cu$_{0.99}$Zn$_{0.01}$O$_{3}$. (b) C$_p$/T of Sr$_{2}$Cu$_{0.99}$Zn$_{0.01}$O$_{3}$ plotted as a function of T under various applied magnetic fields} \label{TN_Zn_H} \end{figure}

In Fig. \ref{Pure_FD}(a) and \ref{Pure_FD}(b) temperature dependence of specific heat for the nominally pure crystal is shown for various values of applied magnetic fields up to 90 kOe. The data are shown as C$_P$/T plotted against T for H $\parallel$ \textit{a} and H $\parallel$ \textit{b}. We found that in both crystal orientations, applied magnetic field has a rather unusual effect. With increasing field the anomaly progressively shifts to higher temperatures, and simultaneously its size gradually increases . In both crystal orientations, the rate of increase of T$_N$ is found to be the same ($\sim$ 1.5 \% per Tesla) over the whole range as shown in Fig. \ref{TN_Zn_H}(a). 

At lower temperatures the specific heat data for the two orientations differ slightly even for the zero-field measurements. Since our measurements along both the orientations were performed on same crystal specimen, the minor differences at low-temperatures, in particular in the zero-field data, are not intrinsic. From previous investigations it is known that in insulating magnetic crystals (e.g., the spin ice compound Dy$_2$Ti$_2$O$_7$ \cite{Snyder2001} and several low-dimensional quantum magnets \cite{Sahling2016}), the slow relaxation of low-energy excitations can introduce some dependence of the measured specific heat on extrinsic parameters, such as, the measurement duration, thermal anchoring of the sample, applied magnetic field, etc. \cite{Lasjaunias1996,Sahling2016}. A combination of these factors might be related to the observed orientation dependence of specific heat at low-temperatures. Since in the region of magnetic transition the specific heat for the two orientations nearly overlaps, the increasing behavior of T$_N$ with field shown in Fig. \ref{TN_Zn_H}(a) is independent of the minor uncertainties in the specific heat at low-temperatures.

C$_p$/T vs. T plots of Zn-doped Sr$_2$CuO$_3$ for H $\parallel$ \textit{a} are shown in Fig. \ref{TN_Zn_H} (b). Similar plots were obtained for H $\parallel$ \textit{b} (data not shown).Variation of T$_N$ with applied field in this case too showed a linear \textit{isotropic} increase (see, Fig. \ref{TN_Zn_H}(a)) at a rate of $\sim$1.4\% per Tesla, which is nearly the same as for the undoped crystal. This behavior is rather intriguing because in conventional antiferromagnets applied magnetic field tends to broaden and lower the magnetic transition. In quasi-one-dimensional antiferromagnets it has however been argued that the primary effect of applied field is to increase the spin-correlation length which in turn increases T$_N$ [Refs. \citenum{Blume1975,Lovesey1979}]. In several Mn (spin-5/2) based quasi-one-dimensional antiferromagnets, T$_N$ was reported to increase significantly in the presence of a magnetic field \cite{Jonge1978}. For example, in the spin-5/2 chain compound tetramethylammonium-manganese trichloride (commonly referred to as TMMC),  T$_N$ increases by almost 20 \% under a modest field of 10 kOe \cite{Dupas1976}.

In the spin-1/2 chains, the presence of strong quantum fluctuations suppress the development of long-range spin ordering. Imry et al. looked at the effect of magnetic field on T$_N$ by taking into account the quantum fluctuations in their semi-classical calculations \cite{Imry1975}. They found that the primary effect of magnetic field is to suppress the quantum fluctuations which in-turn results in a gradual increase of T$_N$ with increasing magnetic field. They estimated that the increase of T$_N$ in the spin-1/2 chain would be almost two orders of magnitude smaller than without quantum fluctuations \cite{Borsa1978}. The behavior of pure and Zn-doped Sr$_2$CuO$_3$ qualitatively agrees with these arguments because the ordering temperature increases only gradually and not as rapidly as in the classical spin chains (TMMC, for example) . The observed increase in the size of the specific heat anomaly with increasing applied magnetic field is also in line with their prediction \cite{Imry1975}. It is shown that the jump $\Delta$C$_p$ in the specific heat will be reduced to almost 50 \% of the classical value in the presence of quantum fluctuations. Hence, suppression of quantum fluctuations under applied magnetic field should increase the size of the specific heat anomaly as is found to be the case experimentally. In the Ni-doped crystal where no long-range magnetic ordering was detected under zero-field condition, application of a magnetic field as high as 90 kOe only change the specific heat marginally at low temperatures but does not induce any magnetic ordering.      

\subsection{Schottky contribution due to  odd-length segments}
\begin{figure} \centering \includegraphics[width=4.2cm]{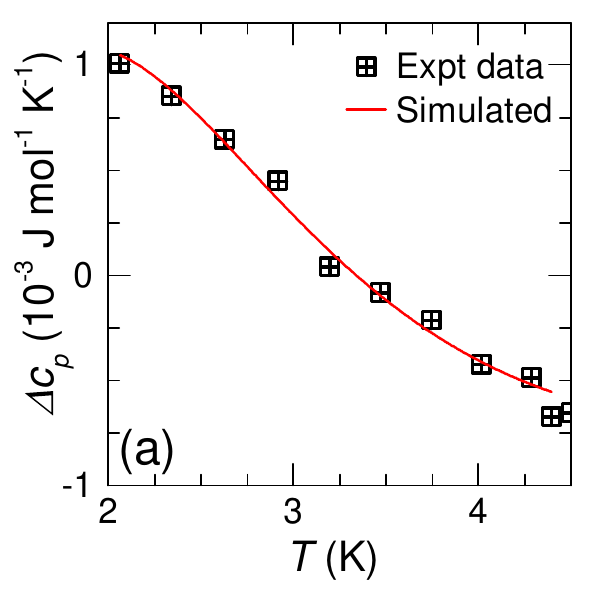}\includegraphics[width=4.2cm]{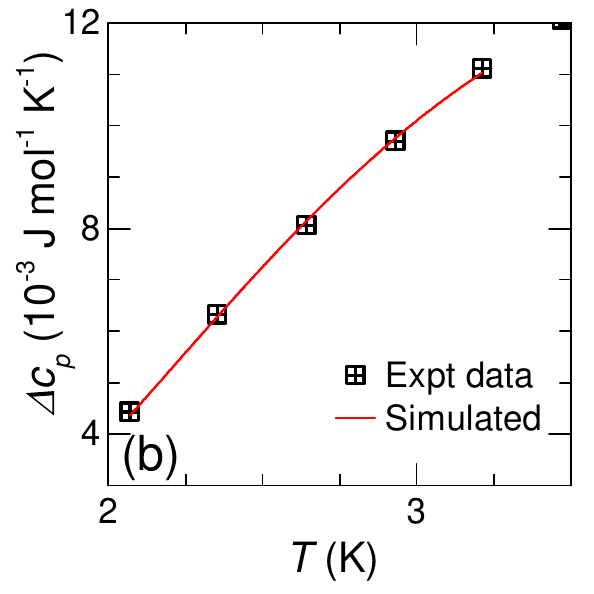} \caption{(a) Specific heat difference $\Delta$C$_p$ obtained by subtracting C$_p$ of Sr$_{2}$CuO$_{3}$ measured under fields of 70 and 90 kOe. (b) Specific heat  difference $\Delta$C$_p$ obtained by subtracting C$_p$ of Sr$_{2}$CuO$_{3}$ from that of  Sr$_{2}$Cu$_{0.99}$Zn$_{0.01}$O$_{3}$ both measured under a field of 90 kOe. The solid lines are the Schottky fits (see text for details)} \label{Pure_Sch} \end{figure}

In a spin chain with an effective chain-break concentration of \textit{p}, the average number of odd- or even-length segments will be \textit{$\sim{p/2}$}. At low-temperatures, the odd-length segments contribute to a Curie-tail to the magnetic susceptibility and a Schottky contribution to the specific heat due to the uncompensated spin-$1/2$. The purpose of this section is to extract the Schottky contribution and hence the effective defect concentration using the specific heat data and check if that is consistent with the value reported previously by analyzing the magnetic susceptibility \cite{Karmakar2015a}. For this purpose, we analyzed the difference specific heat $\Delta$C$_p$ = C$_p$(9) - C$_p$(7) of Sr$_{2}$CuO$_{3}$, where C$_p$(9) and C$_p$(7) are specific heats measured under 90 and 70 kOe, respectively. Using this procedure, the phonon and spinon contributions will be effectively removed \cite{Ramirez1994}. Moreover, since the relaxation times are not expected to vary much between these two  nearby field values, the effect of slow relaxation will be mitigated in $\Delta$C$_p$. We fitted $\Delta$C$_p$ using the Schottky expression taking the defect concentration (\textit{p}) and Zeeman splittings $\Delta$$_7$ and $\Delta$$_9$ under 70 and 90 kOe as the fitting parameters. The fitting was done in the temperature range from T = 2 K up to 4 K. At higher temperatures, the presence of magnetic anomaly does not allow this procedure to be used. A satisfactory fit in this temperature range, as shown by the solid line in Fig. \ref{Pure_Sch}(a), gives the following values of the fitting parameters: \textit{p} = 0.0039(1) per Cu, $\Delta$$_7$ = 7.5 K and $\Delta$$_9$ = 8.5 K. The value of \textit{p} from the susceptibility analysis is about 0.0047 per Cu \cite{Karmakar2015a}, which reflects a good agreement between the two techniques. The value of $\Delta$$_7$ and $\Delta$$_9$ for a free spin-$1/2$ (g-factor = 2) comes out to be $\approx9$ K and 12 K, respectively. The fitted values are somewhat smaller but not unreasonable given that the uncompensated spins on the odd-length segments of the spin-1/2 chain are not free to align isotropically because of the interchain coupling. 

In the Zn-doped case where the spin-$1/2$ concentration is higher, a similar analysis could not be applied because of the lower ordering temperature, which leaves an insufficiently narrow temperature range over which to perform the fit. However, a quick estimate was made by fitting the difference specific heat obtained by subtracting the specific heat of undoped crystal under 90 kOe from that of the Zn-doped crystal measured under the same field. The fitting of the subtracted data (shown in Fig. \ref{Pure_Sch}(b)) gives the ‘excess’ effective chain break concentration, i.e., due to doped Zn$^{2+}$ alone. The value obtained from the fit is \textit{p} = 0.0072(1) Zn per Cu, which compares favorably with the value of 0.006 from the magnetic susceptibility analysis. 

\section{Summary and Conclusions}

The main thrust of this paper is to investigate if the spin gap opening recently reported due to Ni-doping in SrCuO$_2$ \cite{Simutis2013} also exists for Zn-doping. This question is relevant because being a spin-0 defect, Zn-doping is expected to break the chain into finite length segments analogous to the Ni$^{2+}$ (spin-1) case. Unexpectedly, we found no evidence of spin gap with Zn$^{2+}$ doping. We believe that this difference is due to the spatial size of the defect. In Ni$^{2+}$ case, the chain breaks at the impurity site due to the Kondo-singlet, which magnetically isolates the adjoining chain segments. In the Zn$^{2+}$ case, however, the defect is localized over one lattice site and the chain segments in this case are bridged due to appreciable second-nearest-neighbour coupling. 
The second goal of this work was to examine if a similar spin gap is induced in other spin-1/2 chains. For this purpose we examined 1 \% Ni-doped Sr$_2$CuO$_3$ and found the low-lying excitations to be gapped, as was found earlier for Ni-doped SrCuO$_2$ \cite{Simutis2013}. 
Our third goal was to investigate the magnetic ordering of doped and undoped chains. For this purpose we studied the specific heat of pure, Zn-doped and Ni-doped Sr$_2$CuO$_3$ crystals. We found that doping with Zn$^{2+}$ suppresses the transition temperature in accordance with the previous theoretical calculation \cite{Eggert2002}. However, suppression with Ni$^{2+}$ doping was found to be more severe than expected. The effect of magnetic field on the magnetic transition temperature and the size of the associated magnetic anomaly were investigated. We found that magnetic field increases the magnetic ordering temperature and augments the size of the magnetic anomaly. Both these results are in qualitative agreement with a semi-classical theory which shows that suppression of quantum fluctuations in quantum spin-1/2 chains should enhance the size of the specific heat anomaly and the magnetic ordering temperature \cite{Imry1975}. Low-temperature specific heats of the nominally pure and Zn-doped crystals were analyzed to extract the effective defect concentrations. These values were found to be in good agreement with our previous estimations using the magnetic susceptibility analysis \cite{Karmakar2015a}.  
\\

\textbf{\textit{Note added:}} While this paper was being written another report by Simutis et al. appeared (arXiv:1611.04183v1) where INS for 1 \% and 2 \% Ni-doped Sr$_2$CuO$_3$ were reported. Our data shown in Fig. 1b are in good agreement with the data shown in Fig. 2b of this report. 
\\

\begin{acknowledgments}
	We acknowledge financial support for travelling under the Indo-Swiss research grant no. INT/SWISS/ISJRP/PEP/P-06/2012. We thank Bertran Roessli, Uwe Stuhr and Amy Poole for their help in carrying out the neutron scattering experiments. We are thankful to Yannic Utz and Hans-Joachim Grafe for useful discussions.     
\end{acknowledgments}

\bibliography{HC_213}
\bibliographystyle{apsrev4-1}

\end{document}